\documentclass[preprint,12pt]{elsarticle}

\usepackage[T1]{fontenc}
\usepackage[utf8]{inputenc}
\usepackage{lmodern}
\usepackage{microtype}
\usepackage{amsmath,amssymb}
\usepackage{siunitx}
\usepackage{booktabs}
\usepackage{graphicx}
\usepackage{xcolor}
\usepackage{xurl}
\usepackage[hidelinks]{hyperref}
\journal{Scripta Materialia}
\biboptions{sort&compress}
\DeclareSIUnit\angstrom{\text{\AA}}
\graphicspath{{figures/}}

\newcommand{\aSiN}{\textit{a}-Si$_3$N$_4$}

\title{The role of density in shaping electron trapping in amorphous silicon nitride}

\author[inst1,inst2]{Jonathon Cottom}
\author[inst1,inst2]{Emilia Olsson\corref{cor1}}

\affiliation[inst1]{organization={Institute for Theoretical Physics, University of Amsterdam},
            addressline={Science Park 904},
            city={Amsterdam},
            postcode={1098 XH},
            country={the Netherlands}}
\affiliation[inst2]{organization={Advanced Research Center for Nanolithography},
            addressline={Science Park 106},
            city={Amsterdam},
            postcode={1098 XG},
            country={the Netherlands}}
\cortext[cor1]{Corresponding author.}

\begin{document}
\begin{frontmatter}
\ead{k.i.e.olsson@uva.nl}

\begin{abstract}
The electronic properties of amorphous silicon nitride vary with processing, which often changes stoichiometry and density together. Separating their contributions is essential to connect film structure with electron trapping. Across 499 stoichiometric configurations spanning 2.86--3.19~g\,cm$^{-3}$, porosity controls the trapping mixture. Lower density increases accessible void fraction from 8.4 to 12.6\%, expands internal surface and more than triples the supply of free three-coordinate Si precursors. Capture at pre-existing Si dangling bonds (K centres) rises from 9 to 29\%, while induced-K and polaronic trapping decline. A neutral precursor fingerprint predicts the trapping mixture in held-out configurations and reproduces its density dependence. Local compensation controls K-centre activation, electronic competition selects capture, and bond strain predicts polaronic relaxation or induced-K formation. Routes involve distinct network volumes but retain their characteristic depths and relaxation energies across the density range. At fixed stoichiometry, density controls the trapping mixture through porosity and precursor availability.
\end{abstract}


\end{frontmatter}

\section{Introduction}

Amorphous silicon nitride (\aSiN{}) combines mechanical resilience with excellent dielectric performance, underpinning its widespread use in protective coatings, microelectromechanical systems, photovoltaics, and semiconductor manufacturing \cite{Yen2003Protective,Bhatt2009MEMS,Ren2010Trapping,Parsons1991LowHydrogen}. Its capacity for long-term retention of injected charge gives this protective dielectric an active electronic role as the storage medium in modern charge-trap memories \cite{Lina2008,Shaposhnikov2007,Kim2023Review}. Because this charge retention relies on localized trap states~\cite{Kim2025Spectroscopy}, memory performance is fundamentally tied to the structure established during deposition and subsequent annealing or densification steps\cite{Parsons1991LowHydrogen,Beliaev2022Optical,Cottom2024Jekyll}. Determining whether these structural differences reflect altered local bonding or internal porosity is therefore central to connecting film processing with the environments that capture and retain charge.

Magnetic-resonance measurements identify Si- and N-centred dangling bonds and directly link the Si dangling bond (K-centre) to charge trapping \cite{Krick1988Dominant,Warren1990ENDOR,Warren1990a}. Electrical and optical measurements also support Si--Si trapping configurations \cite{Gritsenko2016Nature,Gritsenko2026NegativeCorrelation}. Deposition and thermal treatment alter film density, trap concentration, and the balance of shallow and deep states \cite{Beliaev2022Optical,Ren2010Trapping,Kim2006TrapDistribution}, with direct consequences for capture, transport and charge retention \cite{Liao2013Trapping,Arreghini2008Retention}. Density functional theory (DFT) calculations have connected electronic structure to defect chemistry, passivation and network morphology \cite{Kang2018NegativeU,Hueckmann2024Intrinsic, Cottom2024Jekyll,Jarolimek2010Atomistic,Vedula2012Topological}, and established polaronic, induced-K and Si--Si trapping routes \cite{Hueckmann2024Intrinsic,Wilhelmer2024Polaron, Wilhelmer2026ElectronTrapping}. Processing changes composition and density together; their contributions to the trapping mixture must therefore be separated. Establishing how density changes trapping at fixed stoichiometry is essential to connect film structure with charge-storage behaviour.

Connecting density to electron trapping requires identifying the precursor environments available in the neutral network, their electronic activity, and their competition for charge~\cite{Hueckmann2026DryOxidation}. A relaxed trap establishes an outcome, but does not reveal why that environment captured the electron rather than the alternatives present. The trapping mixture therefore depends on both how frequently viable precursors occur and how they respond during capture and relaxation. Resolving these contributions requires structural and electronic descriptors evaluated before charging and tested against the resulting trapped states. Testing their predictive power across the density range distinguishes changes in precursor supply from changes in the local trapping response. This provides the atomistic basis for a quantitative structure--property relation from which testable design rules can be derived.

Here we establish this relation using stoichiometric \aSiN{} configurations spanning 2.86--3.19~g\,cm$^{-3}$ \cite{Hueckmann2024Intrinsic,Cottom2026Vitriflow}. The local Si--N network changes little; the density contrast is expressed principally through nanoscale porosity, with greater accessible void volume and internal surface at lower density. That additional surface more than triples the supply of free three-coordinate Si precursors and shifts trapping from induced-K and polaronic responses towards capture at pre-existing K centres. A structural precursor fingerprint predicts the trapping mixture across the density range. Local compensation determines whether a K-centre precursor is electronically active, state localisation and energetic ordering determine whether it captures, and neutral bond strain predicts polaronic relaxation or induced-K formation. The resulting routes recruit distinct volumes of the network. Density therefore changes precursor supply and the trapping mixture at fixed stoichiometry, while the routes retain their characteristic depths and relaxation energies.

\section{Methods}

We analysed 499 stoichiometric Si$_{120}$N$_{160}$ configurations: 415 from Ref.~\cite{Cottom2026Vitriflow} and 84 from Ref.~\cite{Hueckmann2024Intrinsic}, retained after electronic-consistency screening (Supplementary Section~S1.5). Both sources used MG2 $NpT$ melt--quench simulations followed by HSE06 refinement.\cite{Marian2000MG2} Their common density--porosity relation and route statistics over the overlapping density interval support analysis as one population spanning 2.86--3.19~g\,cm$^{-3}$ (Supplementary Fig.~S1). The three-route analysis contains 496 cells; three N-centred traps are treated separately. Si--Si formation occurs in three of the routed cells and is flagged within their assigned routes. Generation and structural validation are described in the original studies.

Spin-polarised DFT calculations were performed with CP2K/Quickstep~\cite{CP2K} using HSE06 \cite{HSE06_original,HSE06_erratum,Krukau2006HSE06} with the auxiliary density matrix method \cite{ADMM_2010}, DZVP-SR-MOLOPT basis sets \cite{MOLOPT_Basis_set}, GTH pseudopotentials \cite{GTH1,GTH2,Krack_05}, and auxiliary density cutoffs of 650 and 70~Ry. Neutral structures were optimised before addition of one electron, after which the $q=-1$ structures were geometry optimised at fixed cell. Total energies and eigenvalues include finite-size charge corrections (Supplementary Section~S4). The three Si-centred routes were classified as capture at a pre-existing K centre, induced K-centre (iK) formation through Si--N bond detachment, or topology-preserving polaron formation. Tensile bond strain was characterised using the longest neutral Si--N bond, $d_4^0$, as a geometric proxy.

Void geometry was obtained from a periodic covalent-clearance field on a \SI{0.15}{\angstrom} grid using Cordero radii~\cite{Cordero2008Covalent} and a \SI{0.8}{\angstrom} probe. Localisation was expressed as an effective atom count from molecular-orbital and real-space inverse participation ratios; mobility edges use participation over $N/4$ atoms. State-aligned graph-wavelet fingerprints combine local structural and electronic fields to describe trapping precursors~\cite{cottom2026statealignedgraphwavelet}. A regularised multinomial predictor was fitted to three neutral precursor descriptors on a random half of the routed cells and tested on the other half, repeating the split 200 times. Density-dependent changes are evaluated at 2.92 and 3.15~g\,cm$^{-3}$, the 2.5th and 97.5th density percentiles, with cell-bootstrap 95\% intervals. Descriptor, potential, correction and statistical definitions are given in the Supplementary Methods.

\section{Results and discussion}

\subsection{Porosity controls the supply of trapping precursors}

\begin{figure}[hbt!]
\centering
\includegraphics[width=\textwidth,keepaspectratio]{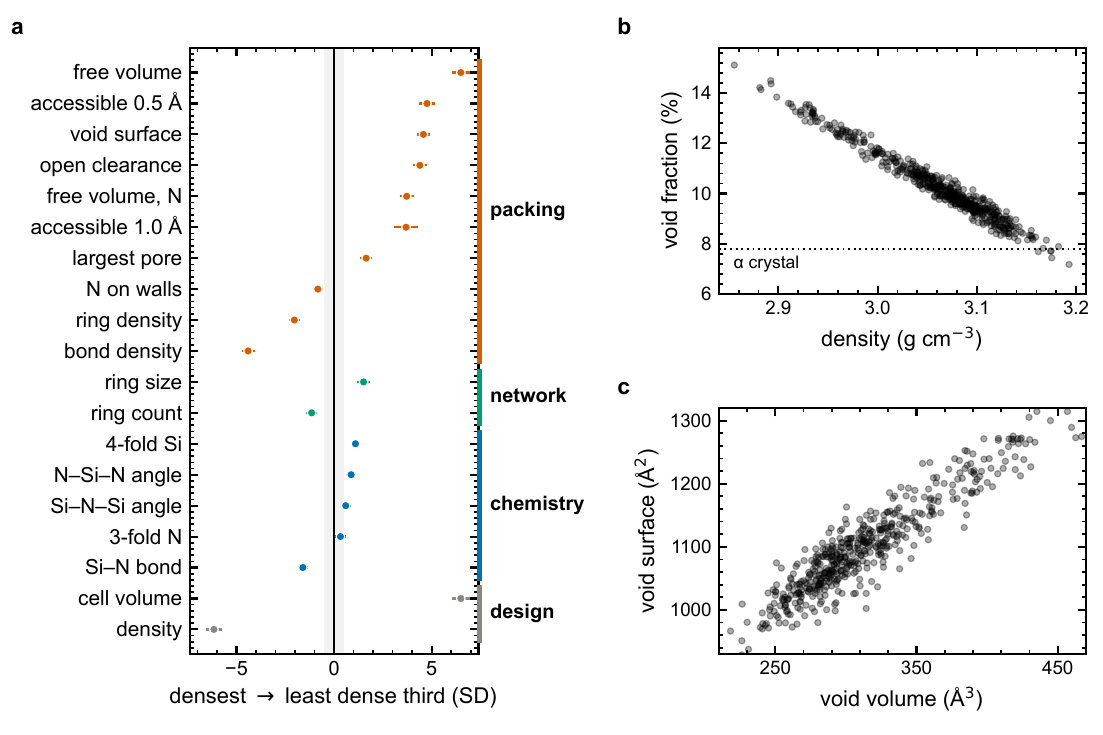}
\caption{\textbf{Density varies through packing and void space.} (a) Descriptor differences between the least dense and densest thirds of the 499 configurations, normalised by the standard deviation of the densest third; error bars are 95\% intervals. (b) Accessible void fraction against density for a \SI{0.8}{\angstrom} probe; the dotted line is crystalline $\alpha$-Si$_3$N$_4$ evaluated with the same measure. (c) Accessible-void surface area against void volume.}
\label{fig:structure}
\end{figure}

The density contrast is expressed through porosity rather than a change in the local Si--N network. The mean Si--N bond length changes by only \SI{0.01}{\angstrom} across the density range and the angle distributions remain closely matched, while packing and void descriptors show much larger shifts (Fig.~\ref{fig:structure}a; Supplementary Fig.~S2). As density decreases from 3.15 to 2.92~g\,cm$^{-3}$, accessible void fraction rises from 8.4 to 12.6\% and internal surface from 997 to \SI{1225}{\angstrom\squared} per cell (Fig.~\ref{fig:structure}b,c). Crystalline $\alpha$-Si$_3$N$_4$ gives 7.8\% by the same void measure. Specific internal surface increases from 337 to \SI{389}{\angstrom\squared} per \SI{1000}{\angstrom\cubed}, confirming a change in pore morphology beyond the increase in cell volume.

\begin{figure}[hbt!]
\centering
\includegraphics[width=\textwidth,keepaspectratio]{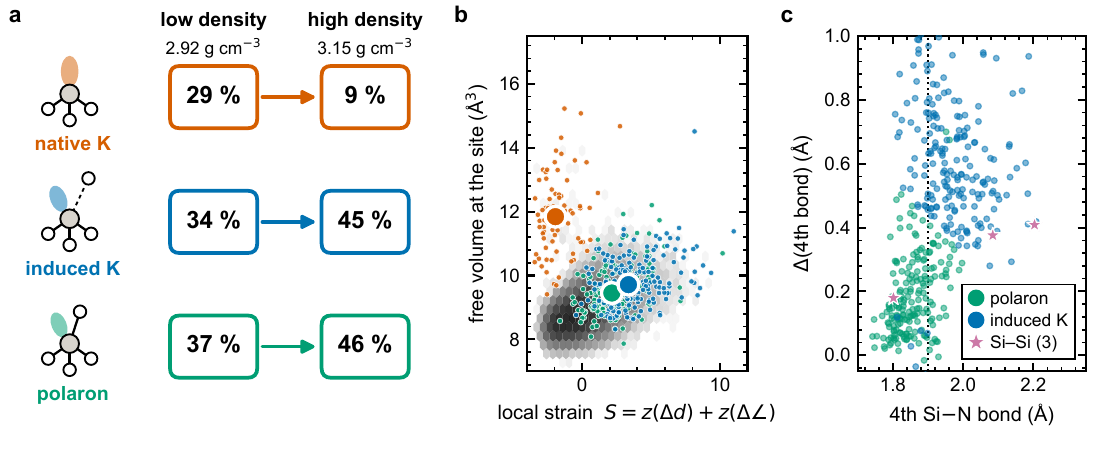}
\caption{\textbf{Porosity changes the trapping mixture and its neutral precursors.} Orange, blue and green denote native K-centre, induced-K and polaronic trapping throughout. (a) The three routes and their fractions estimated from the density trend at 2.92 and 3.15~g\,cm$^{-3}$; 496 cells enter the route analysis. Si is grey and N white. (b) Local free volume and strain, $S=z(\Delta d)+z(\Delta\theta)$, for neutral Si environments; grey denotes candidate sites, colours the eventual trap sites and large symbols route medians. (c) Charged-state extension, $\Delta d_4$, against the neutral length $d_4^0$ of the most extended Si--N bond at polaronic and induced-K sites. The vertical guide marks $d_4^0=\SI{1.90}{\angstrom}$; stars identify three Si--Si reconstructions within the routed population.}
\label{fig:supply}
\end{figure}

This porosity change redistributes the trapping routes (Fig.~\ref{fig:supply}a). Capture at pre-existing K centres rises from 9 to 29\%, while induced-K formation falls from 45 to 34\% and polaronic trapping from 46 to 37\%; Si--Si formation remains rare. The total K-like population increases from 54 to 63\%. Its composition shifts from predominantly induced-K centres to approximately equal mixture of pre-existing and induced centres.

The origin of this shift is already encoded in the neutral structures (Fig.~\ref{fig:supply}b; Supplementary Fig.~S9a). Pre-existing K-centre precursors occupy high-free-volume, comparatively low-strain environments at internal void surfaces; the fraction lying within \SI{2.6}{\angstrom} of a cavity rises from 57 to 79\% with decreasing density. Induced-K precursors occupy high-strain sites in the tensile tail of the neutral Si--N bond distribution, while polarons occupy an intermediate region in which the distorted Si--N shell remains intact. Porosity changes these supplies selectively. The mean number of uncapped three-coordinate Si environments increases from 0.12 to 0.45 per cell, whereas the number of four-coordinate sites with $d_4>\SI{1.90}{\angstrom}$ and the peak local strain remain essentially unchanged. Lower density therefore does not produce uniformly greater disorder; it specifically exposes more dangling-bond precursors at internal void surfaces.

The polaronic and induced-K routes follow a common local response (Fig.~\ref{fig:supply}c). Charged-state elongation increases with the initial tensile strain of the fourth Si--N bond, while the other three bonds respond similarly for both endpoints. Relaxation remains polaronic at lower strain and increasingly produces induced K centres towards the upper tail of the neutral bond-strain distribution. The crossover occurs near $d_4^0=\SI{1.90}{\angstrom}$ across the density range, and the rare Si--Si cases lie along the same bond-extension coordinate. Neutral tensile bond strain therefore predicts whether polaronic relaxation arrests or continues to induced-K formation.

\subsection{Local bonding and electronic competition determine which K-centre precursors trap}

\begin{figure}[hbt!]
\centering
\includegraphics[width=\textwidth,keepaspectratio]{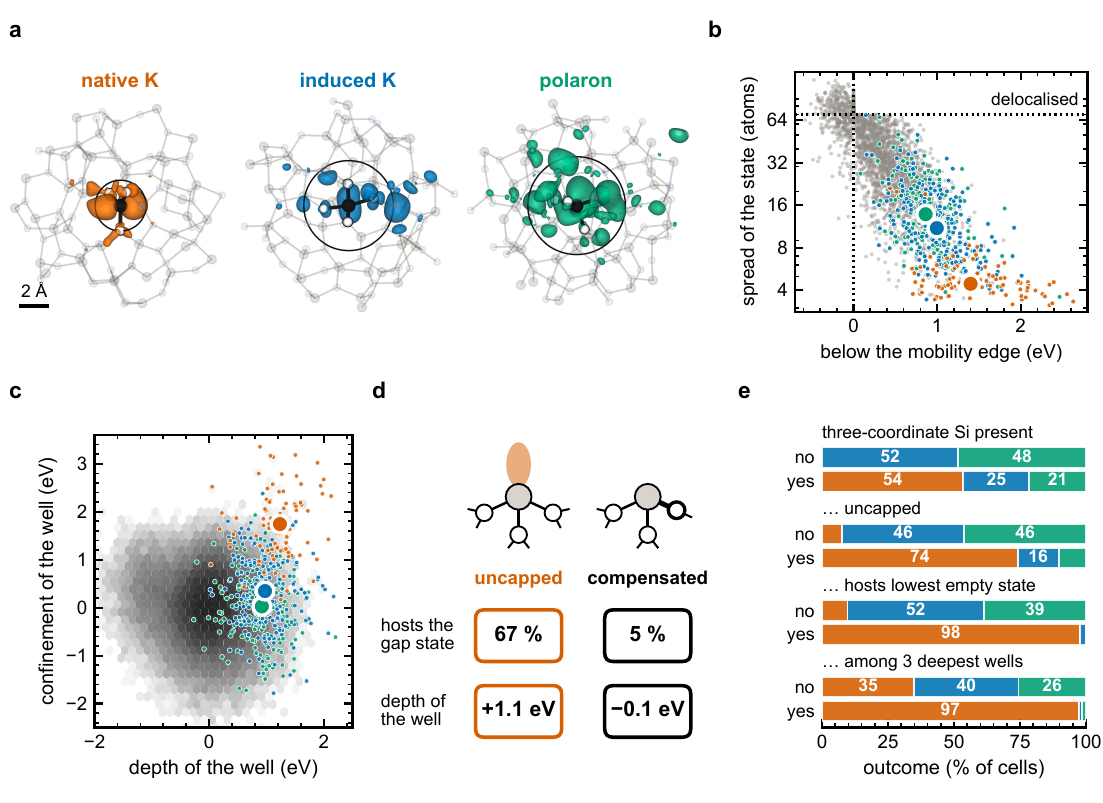}
\caption{\textbf{Neutral electronic structure selects active trapping sites.} (a) Precursor states for one example of each route: isosurfaces enclose half of the added electron at fixed nuclei. The final trap atom is dark, its N neighbours white; circles mark the trap-centred radius enclosing half of the electron. (b) Precursor depth below the mobility edge and effective-atom spread; large symbols are route medians, grey points higher empty states. The mobility edge is the first empty state spread over at least $N/4$ atoms. (c) Depth and confinement of the neutral electrostatic potential well at Si sites; grey denotes all Si sites and colours the eventual traps. Depth is relative to the mean Si-site potential of each cell; confinement is the rise to the surrounding shell. (d) Gap-state hosting and potential-well depth for uncapped and N-compensated three-coordinate Si. (e) Route fractions for cells that do not (upper bars) and do (lower bars) meet each stated condition.}
\label{fig:gap}
\end{figure}

Native K-centre capture begins from a detached, localised neutral LUMO approximately \SI{1.4}{\electronvolt} below the delocalisation onset (Fig.~\ref{fig:gap}a,b). Induced-K and polaronic precursors instead lie near the band edge and extend over more of the network. Pre-existing K-centre capture therefore occupies an already formed gap state, whereas induced-K and polaronic trapping develop from broader band-tail states during relaxation. This electronic preference is also resolved in the neutral electrostatic potential: the eventual trapping atom lies among the five deepest Si-site wells in 65\% of cells, and pre-existing K-centre precursors occupy the most confined wells (Fig.~\ref{fig:gap}c).

Three-coordinate Si does not automatically supply this state. When locally compensated by a bonded two-coordinate N, only 5\% of sites host the detached state, compared with 67\% of uncapped sites (Fig.~\ref{fig:gap}d). The compensating Si--N bond is approximately \SI{0.09}{\angstrom} shorter than an ordinary Si--N bond and is the shortest at the Si site in 91\% of cases (Supplementary Fig.~S6a). Cells containing only compensated candidates accordingly recover the ordinary band-edge response: their LUMOs extend over approximately 10--13 atoms, closely matching cells with no dangling-bond candidate, whereas a free candidate produces a detached LUMO localised over approximately five atoms (Supplementary Fig.~S6b). Compensation also reduces the potential-well depth from +1.1 to \SI{-0.1}{\electronvolt} (Fig.~\ref{fig:gap}d). Coordination identifies a candidate; local bonding determines its electronic activity.

Activation still does not guarantee capture. Only 74\% of cells containing an active uncapped candidate trap through the pre-existing K-centre route; in approximately one cell in four, induced-K or polaronic trapping occurs elsewhere (Fig.~\ref{fig:gap}e). The neutral empty-state manifold resolves this competition (Fig.~\ref{fig:gap}e; Supplementary Fig.~S6c). In the candidate-resolved analysis, captured states are compact, extending over approximately four atoms, and form the cell LUMO. Bypassed candidates can lie deeper below the mobility edge, but remain distributed over approximately twelve atoms and are not the lowest empty state; a more localised competing state captures instead. When an uncapped site hosts the lowest empty state, K-centre capture occurs in 98\% of cells (Fig.~\ref{fig:gap}e). Capture is therefore determined by localisation and energetic ordering together, not state depth alone.

The neutral network consequently provides a prospective three-stage prediction: undercoordination supplies a candidate, local compensation determines whether it supports a detached K-like state, and competition among the neutral empty states determines whether that state captures. The source populations reproduce this selection hierarchy (Supplementary Fig.~S1f). The increase in pre-existing K-centre trapping therefore arises from the greater supply of viable void-surface precursors, rather than from a change in the electronic criterion for capture.

\subsection{Trap energetics and prediction of the trapping mixture}

\begin{figure}[hbt!]
\centering
\includegraphics[width=\textwidth,keepaspectratio]{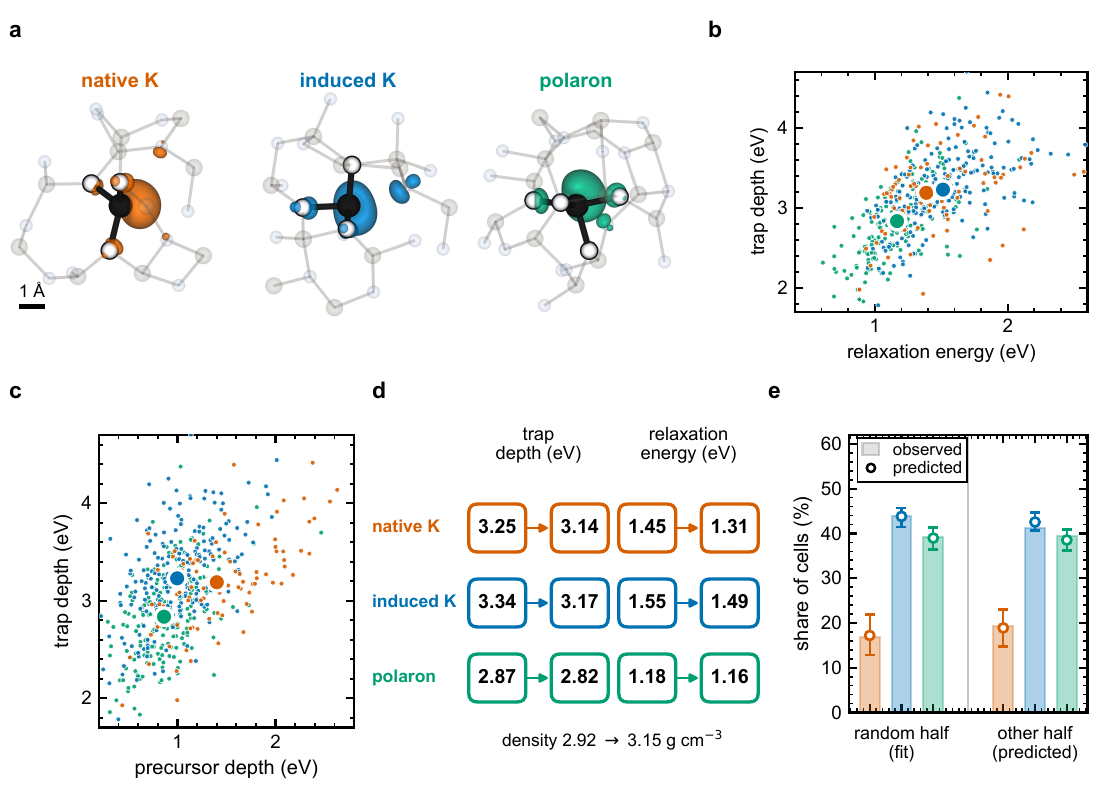}
\caption{\textbf{Relaxation establishes the characteristic trap energies.} (a) Relaxed states of the example cells in Fig.~\ref{fig:gap}a; each isosurface encloses half of the electron. (b) Finite-size-corrected relaxation energy and trap depth for routed cells; large symbols are route medians. Relaxation energy is the positive energy released between the vertical and relaxed $q=-1$ endpoints. (c) Corrected relaxed trap depth against neutral precursor depth. (d) Estimated trap depth and relaxation energy within each route at 2.92 and 3.15~g\,cm$^{-3}$. (e) Observed fractions (bars) and mean predicted probabilities (circles) for a random training half and the held-out half; error bars are cell-bootstrap 95\% intervals. Density-resolved held-out predictions and repeated-split performance are given in Supplementary Fig.~S9.}
\label{fig:relax}
\end{figure}

Immediately after electron addition, the eventual trap atom already carries the largest spin population in 82\% of cells. Relocation occurs mainly when the vertical state is distributed across competing regions. The subsequent ionic response distinguishes the routes (Supplementary Fig.~S8c). Capture at a pre-existing K centre is compact: the trap atom moves by \SI{0.39}{\angstrom}, but its bonded neighbours by only \SI{0.06}{\angstrom}. Polaronic relaxation involves approximately 93 atoms, and induced-K formation extends furthest through the surrounding network. For these two routes, the trap atom moves by 0.25--\SI{0.28}{\angstrom} and its bonded neighbours by \SI{0.15}{\angstrom}. A large displacement of the trapping atom therefore need not imply an extended network response.

Relaxation then creates the final localised state (Fig.~\ref{fig:relax}a; Supplementary Fig.~S8a). Pre-existing K-centre states change little in extent because they are already localised in the neutral network. Induced-K and polaronic precursors can extend over tens of atoms, but both collapse to final states involving approximately five effective atoms; the independent molecular-orbital IPR reproduces the same ordering. Similar final localisation therefore conceals different origins: K-centre capture occupies an existing localised state, whereas induced-K and polaronic trapping form their compact states through network relaxation. The accompanying energetic stabilisation is substantial (Fig.~\ref{fig:relax}b,c). Corrected charged-state relaxation releases approximately 1.4, 1.5 and \SI{1.2}{\electronvolt} for pre-existing K-centre, induced-K and polaronic trapping, respectively. The corrected traps lie approximately 3.25, 3.34 and \SI{2.87}{\electronvolt} below the mobility edge, following a precursor-to-trap deepening of about \SI{2}{\electronvolt}. Across the density interval, each route retains its characteristic energy scale (Fig.~\ref{fig:relax}d): induced-K states remain electronically K-like, while polarons remain approximately \SI{0.4}{\electronvolt} shallower.

The neutral precursor fingerprint predicts the resulting mixture. A multinomial model was fitted to a random half of the configurations using uncapped three-coordinate Si, sites in the tensile-strain tail and peak local strain, then applied unchanged to the held-out half (Fig.~\ref{fig:relax}e). Across 200 random splits, it assigns 60\% of held-out routes correctly, compared with 41\% from route fractions alone. It also reproduces the density dependence: as density decreases, K-centre capture rises from 10 to 30\% in prediction and from 9 to 29\% in the fitted observations (Supplementary Fig.~S9). Adding void fraction does not improve held-out prediction. Porosity acts through the precursor environments it supplies; the fingerprint connects these environments to the trapping mixture without a density label. At fixed stoichiometry, density changes the balance of trapping routes while retaining their characteristic energetic response.

\section{Conclusions}

The density contrast in stoichiometric \aSiN{} is expressed primarily through porosity, not a change in the local Si--N network. Lower density increases accessible void volume and internal surface, more than tripling the supply of free three-coordinate Si precursors and increasing capture at pre-existing K centres from 9 to 29\%. Induced-K and polaronic trapping become less frequent, while the three routes recruit distinct volumes of the network: compact K-centre capture, collective polaronic relaxation and the extended response accompanying induced-K formation. This change is predicted from the neutral structure. A precursor fingerprint fitted to half of the configurations reproduces the trapping mixture and its density dependence in the held-out half. Local Si--N compensation determines whether a three-coordinate Si hosts an active gap state, electronic localisation and state ordering determine whether it captures, and neutral tensile bond strain predicts whether relaxation remains polaronic or proceeds to induced-K formation. At fixed stoichiometry, density therefore governs electron trapping by changing pore morphology and the population of viable precursors. The local rules governing activation, capture and relaxation remain common across the density range, and the routes retain their characteristic depths and relaxation energies. Density must therefore be distinguished from composition when interpreting changes in trap populations.


\section*{Declaration of competing interest}
The authors declare that they have no known competing financial interests or personal relationships that could have appeared to influence the work reported in this paper.

\section*{Data availability}
The configurations are publicly available on Zenodo: those from Ref.~\cite{Hueckmann2024Intrinsic} at \url{https://doi.org/10.5281/zenodo.8239085}, and those from Ref.~\cite{Cottom2026Vitriflow} at \url{https://doi.org/10.5281/zenodo.21111823}. The analysis package for this work and frozen versions of the code are archived on Zenodo at \url{https://doi.org/10.5281/zenodo.23235037}. Development code is available on GitHub for graph-wavelet fingerprinting (\url{https://github.com/Olsson-Materials-Modelling/sa-gw}), finite-size corrections (\url{https://github.com/Olsson-Materials-Modelling/cp2k-charge-corrections}) and structural analysis (\url{https://github.com/Olsson-Materials-Modelling/vfa}).

\section*{Acknowledgements}
This work was conducted at the Advanced Research Center for Nanolithography, a public-private partnership between the University of Amsterdam (UvA), Vrije Universiteit Amsterdam (VU), Rijksuniversiteit Groningen (RUG), the Netherlands Organization for Scientific Research (NWO), and the semiconductor equipment manufacturer ASML. This work made use of the Dutch national e-infrastructure with the support of the SURF Cooperative using grant no. EINF-16595 and EINF-8981. The authors thank SURF (www.surf.nl) for the support in using the National Supercomputer Snellius. This work made use of the ARCNL Minerva cluster hosted at HOPSTER at UvA. E.O. is grateful for a WISE Fellowship from the NWO and acknowledges support via Holland High Tech through a public-private partnership grant in research and development within the Dutch top sector of High-Tech Systems and Materials (HTSM).

\bibliographystyle{elsarticle-num}
\bibliography{references}

\end{document}


\begin{center}
{\large\bfseries Supplementary information}\par\vspace{5mm}
{\Large\bfseries The role of density in shaping electron trapping in amorphous silicon nitride}\par\vspace{5mm}
Jonathon Cottom$^{1,2}$ and Emilia Olsson$^{1,2,*}$\par\vspace{3mm}
{\small $^1$Institute for Theoretical Physics, University of Amsterdam, Science Park 904, 1098 XH Amsterdam, the Netherlands\par
$^2$Advanced Research Center for Nanolithography, Science Park 106, 1098 XG Amsterdam, the Netherlands\par
$^*$Correspondence: \href{mailto:k.i.e.olsson@uva.nl}{k.i.e.olsson@uva.nl}}
\end{center}
\vspace{4mm}

\section{Model generation, electronic calculations and analysis populations}
\label{sec:models}

\subsection{Common model lineage}
The configurations were taken from H\"uckmann \emph{et al.}~\cite{Hueckmann2024Intrinsic} and the HSE06 branch of the Vitriflow study~\cite{Cottom2026Vitriflow}. Both sources comprise periodic, stoichiometric Si$_{120}$N$_{160}$ cells generated by MG2 melt-quench simulations and refined using density functional theory.

\subsection{\texorpdfstring{Configurations from Ref.~\cite{Hueckmann2024Intrinsic}}{Configurations from H\"uckmann et al.}}
H\"uckmann \emph{et al.}\cite{Hueckmann2024Intrinsic} generated 280-atom $\beta$-Si$_3$N$_4$-derived cells using MG2/LAMMPS melt-quench simulations with Nos\'e-Hoover $NpT$ dynamics. Their timestep was \SI{0.5}{\femto\second}; thermostat and barostat coupling times were $100\Delta t$ and $1000\Delta t$; melt temperature was \SI{4000}{\kelvin}. Tests covered 112, 280, 336, 672 and 1512 atoms and cooling rates of 0.1, 1 and \SI{10}{\kelvin\per\pico\second}. The selected production size and rate were 280 atoms and \SI{1}{\kelvin\per\pico\second}.

One hundred structures were selected from a 300-cell, PBE-refined parent ensemble by matching its distributions of density, undercoordination, bulk modulus and four-membered-ring content. Trial subsets were compared with fitted parent distributions, and members were replaced iteratively to improve agreement. The selected structures were subsequently refined with HSE06.\cite{Hueckmann2024Intrinsic}

\subsection{\texorpdfstring{Configurations from Ref.~\cite{Cottom2026Vitriflow}}{Configurations from Cottom et al.}}
The Vitriflow source population contains 455 MG2 production cells. The autotune calibration used 20 replicas, a liquid onset of \SI{3500}{\kelvin}, a production temperature of \SI{4000}{\kelvin} and a liquid hold of \SI{107.8}{\pico\second}. Cooling proceeded at \SI{1}{\kelvin\per\pico\second} to \SI{300}{\kelvin}, followed by \SI{100}{\pico\second} of equilibration. Periodic isotropic $NpT$ dynamics used zero target pressure, a \SI{1}{\femto\second} timestep and thermostat/barostat coupling times of 0.50/\SI{5.0}{\pico\second}. PBE and HSE06 refinements were separate descendants of these MG2 parents; the present work uses the HSE06 branch. Coordination-outlier labels in the source study were advisory, so all 455 cells were retained in that study. The electronic-consistency screen applied here is described in Section~\ref{sec:population}.\cite{Cottom2026Vitriflow}

\begin{table}[htbp]
\centering
\caption{Classical generation and density functional theory refinement settings for the two model populations. The preparation schedules are described in Sections~S1.2 and S1.3.}
\label{tab:generation}
\small
\begin{tabular}{@{}>{\raggedright\arraybackslash}p{0.40\textwidth}>{\raggedright\arraybackslash}p{0.25\textwidth}>{\raggedright\arraybackslash}p{0.25\textwidth}@{}}
\toprule
Setting & Ref.~\cite{Hueckmann2024Intrinsic} & Ref.~\cite{Cottom2026Vitriflow}\\
\midrule
Atoms; force field & 280; MG2 & 280; MG2\\
Classical ensemble & $NpT$ & $NpT$, isotropic\\
Timestep (fs) & 0.5 & 1.0\\
Thermostat / barostat coupling (ps) & 0.05 / 0.50 & 0.50 / 5.0\\
Cooling rate (K ps$^{-1}$) & 1 & 1\\
HSE06 source population & 100 selected from 300 PBE cells & 455 MG2 descendants\\
Optimisation energy criterion (eV) & $10^{-7}$ & $10^{-7}$\\
Force criterion (eV \AA$^{-1}$) & $5\times10^{-3}$ & $10^{-3}$\\
\bottomrule
\end{tabular}
\end{table}

\subsection{Electronic structure and charge sequence}
Spin-polarised calculations used CP2K/Quickstep\cite{CP2K}, HSE06\cite{HSE06_original,HSE06_erratum,Krukau2006HSE06}, the auxiliary density matrix method\cite{ADMM_2010}, DZVP-SR-MOLOPT basis sets\cite{MOLOPT_Basis_set} and GTH pseudopotentials.\cite{GTH1,GTH2,Krack_05} The auxiliary plane-wave density and relative cutoffs were 650 and 70 Ry. The source structures were optimised using BFGS with the stopping criteria in Table~\ref{tab:generation}. The calculations in Ref.~\cite{Hueckmann2024Intrinsic} used $\Gamma$-point sampling and tested low- and high-spin initial guesses.

With $R^0$ denoting the relaxed neutral geometry. Electron addition at fixed $R^0$ gives the vertical state, followed by ionic relaxation at fixed cell to $R^-$. The three stages are
\begin{equation}
(R^0,q=0)\;\longrightarrow\;(R^0,q=-1)\;\longrightarrow\;(R^-,q=-1).
\label{eq:sequence}
\end{equation}
Electronic states were evaluated separately for the neutral precursor, vertical charged configuration and relaxed charged endpoint. Atom identities were retained through this sequence to connect each trapping region to its neutral precursor. Finite-size charge corrections were applied to both total energies and Kohn-Sham eigenvalues \cite{Freysoldt2009FNV,Kumagai2014FiniteSize,Naik2018CoFFEE}. Spectral comparisons use corrected eigenvalues (Section~\ref{sec:corrected-levels}); fixed-charge relaxation energies use corrected total energies at both charged endpoints (Section~\ref{sec:corrected-energy}).

\subsection{Analysis population}
\label{sec:population}
The analysis uses 415 of the 455 configurations from Ref.~\cite{Cottom2026Vitriflow} and 84 of the 100 from Ref.~\cite{Hueckmann2024Intrinsic}. Electronic-consistency screening removed 40 and 16 cells, respectively, due to unstable wavefunction convergence associated with (near-)degenerate trap states.

\begin{figure}[p]
\centering
\includegraphics[width=\textwidth,height=0.70\textheight,keepaspectratio]{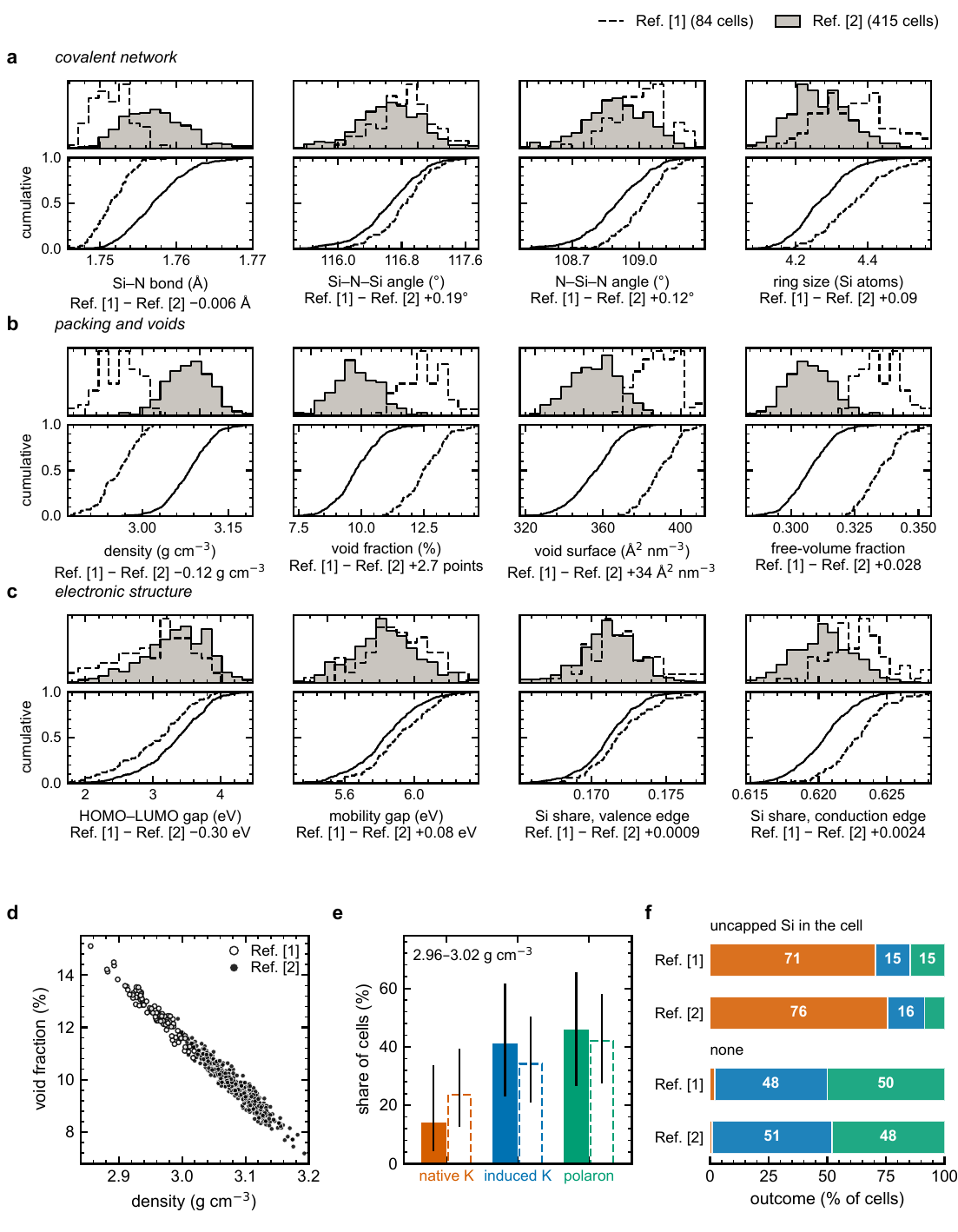}
\caption{\textbf{Source populations and their density dependence.} Solid/filled symbols denote 415 configurations from Ref.~\cite{Cottom2026Vitriflow}; dashed/hollow symbols denote 84 from Ref.~\cite{Hueckmann2024Intrinsic}. (a--c) Per-cell histograms and empirical cumulative distributions of local-network, packing and electronic descriptors. Annotations give the mean for Ref.~\cite{Hueckmann2024Intrinsic} minus that for Ref.~\cite{Cottom2026Vitriflow}. The mobility gap uses the $N/4$ participation threshold. (d) Accessible void fraction against mass density. (e) Route fractions for 22 cells from Ref.~\cite{Cottom2026Vitriflow} and 38 from Ref.~\cite{Hueckmann2024Intrinsic} in the overlapping interval 2.96--\SI{3.02}{\gram\per\centi\metre\cubed}; error bars are 95\% Wilson intervals. (f) Route fractions conditional on the presence or absence of uncapped three-coordinate Si, resolved by source.}
\label{fig:S1}
\end{figure}

The retained 499 cells form one analysis population spanning 2.86--\SI{3.19}{\gram\per\centi\metre\cubed}. The two sources have mean densities of 3.081 and \SI{2.957}{\gram\per\centi\metre\cubed} and overlap over 2.96--\SI{3.02}{\gram\per\centi\metre\cubed}. Figure~\ref{fig:S1} shows their structural and electronic distributions, common density--porosity relation and route fractions in the overlap. 

Of the 499 cells, 496 follow one of the three Si-centred routes: pre-existing K-centre capture, induced-K formation or polaronic trapping. Si--Si formation occurs in three cells. These cells remain in the analysis under their assigned routes and are flagged wherever reconstruction is examined; the three observations are reported individually rather than fitted as a separate class. Table~\ref{tab:routes} gives the complete three-route census and the density-dependent fractions.

\begin{table}[htbp]
\centering
\caption{Si-centred trapping routes. Counts and overall fractions use all 496 routed cells, including those with Si--Si reconstruction. Fractions at 2.92 and \SI{3.15}{\gram\per\centi\metre\cubed} are the fitted density-trend values defined in Section~\ref{sec:stats}, with 95\% cell-bootstrap intervals.}
\label{tab:routes}
\small
\begin{tabular}{@{}lrrr@{}}
\toprule
Route & All cells & $\rho=2.92$ & $\rho=3.15$\\
 & ($n=496$) & \multicolumn{2}{c}{(g\,cm$^{-3}$)}\\
\midrule
Pre-existing K-centre capture & 90 (18.1\%) & 29\% [22,38] & 9\% [4,16]\\
Induced-K formation & 211 (42.5\%) & 34\% [26,43] & 45\% [37,54]\\
Polaronic trapping & 195 (39.3\%) & 37\% [28,46] & 46\% [37,54]\\
\bottomrule
\end{tabular}
\end{table}

\section{Structural descriptors}
\label{sec:structure}

\subsection{Bond graph and local geometry}
\begin{figure}[p]
\centering
\includegraphics[width=\textwidth,height=0.74\textheight,keepaspectratio]{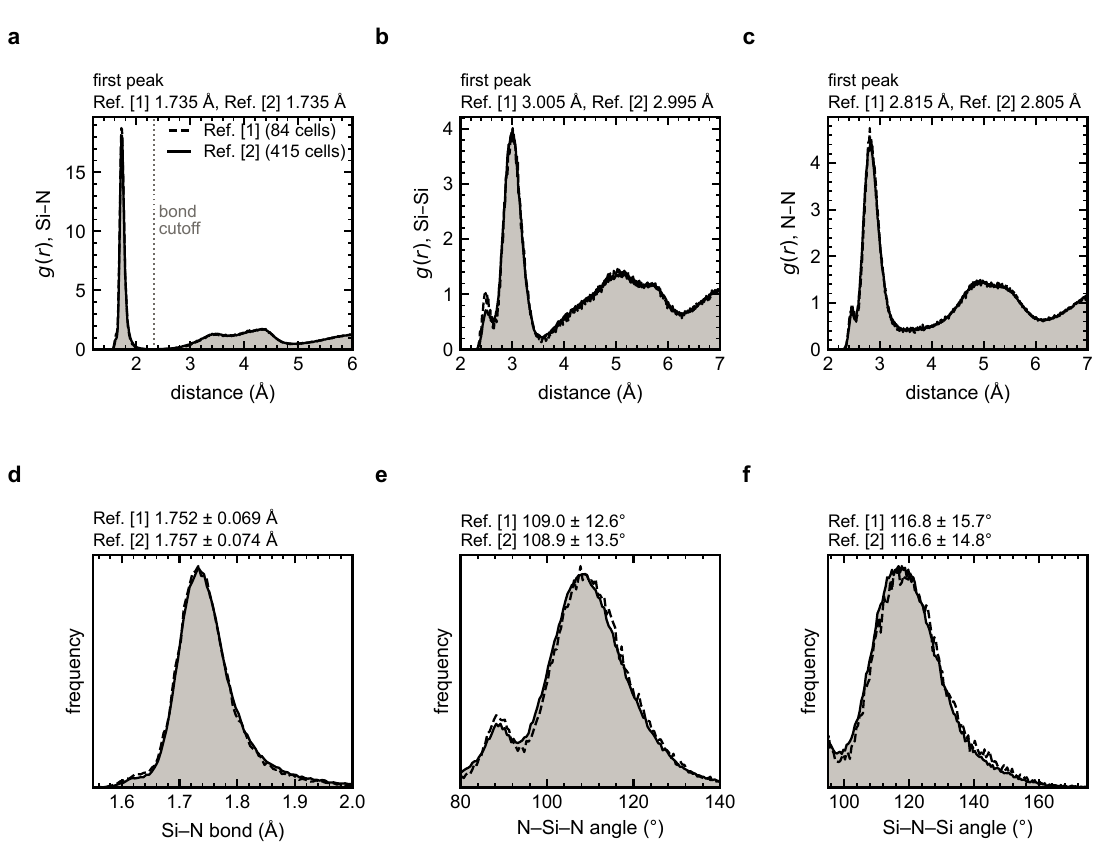}
\caption{\textbf{Local Si--N structure in the two source populations.} Source encoding follows Fig.~\ref{fig:S1}. (a--c) Si--N, Si--Si and N--N radial distribution functions, calculated with equal cell weights and \SI{0.01}{\angstrom} bins using Eq.~\eqref{eq:rdf}. The dotted line in (a) marks the \SI{2.33}{\angstrom} Si--N graph cutoff. (d--f) Si--N distance, N--Si--N angle and Si--N--Si angle distributions over all bonds and angles in each source. Annotations give the means and standard deviations.}
\label{fig:S2}
\end{figure}

Coordination and local geometry were evaluated using a common Si--N cutoff of \SI{2.33}{\angstrom}, corresponding to the first-shell minimum in the pair-correlation analysis. With periodic separation $r_{ij}$,
\begin{equation}
A_{ij}=\ind\{i\in\mathrm{Si},\ j\in\mathrm N,\ r_{ij}\leq\rmin\},
\qquad \rmin=\SI{2.33}{\angstrom}.
\label{eq:graph}
\end{equation}
The reverse edge is implicit. Heteropolar coordination is the number of graph neighbours, with reference coordinations of four for Si and three for N. Histograms in Fig.~\ref{fig:S2} pool bond and angle observations within each source population. Statistical comparisons use the corresponding per-cell summaries. Site fractions were calculated within each cell before ensemble averaging.

Si--N distances and N--Si--N/Si--N--Si angles are calculated from periodic neighbour vectors. For a site with neighbour distances $d_j$ and neighbour-pair angles $\theta_{jk}$, bond-length and angle spreads are
\begin{equation}
\Delta d=\max_jd_j-\min_jd_j,\qquad
\Delta\theta=\max_{j<k}\theta_{jk}-\min_{j<k}\theta_{jk}.
\label{eq:spreads}
\end{equation}

\subsection{Pair correlations and running coordination}
For a radial bin centred at $r$ with width $\Delta r=\SI{0.01}{\angstrom}$, let $n_{\alpha\beta,c}(r)$ be the ordered-pair count for cell $c$. The partial radial distribution function is
\begin{equation}
g_{\alpha\beta,c}(r)=
\frac{n_{\alpha\beta,c}(r)}{N_{\alpha,c}\,4\pi r^2\Delta r\,\rho_{\beta,c}},
\qquad
\rho_{\beta,c}=\begin{cases}
N_{\beta,c}/V_c,&\alpha\ne\beta,\\
(N_{\alpha,c}-1)/V_c,&\alpha=\beta.
\end{cases}
\label{eq:rdf}
\end{equation}
Pair counts were accumulated to \SI{8}{\angstrom} using explicit periodic-image expansion, excluding the zero-distance self term. Each cell contributes equally to the ensemble $g_{\alpha\beta}(r)$. Total and partial distributions use the same ordered-pair convention. The first peak was located from the maximum of the smoothed radial distribution, and the first minimum from the subsequent minimum.

The Si--N running coordination is evaluated directly from the cumulative neighbour count per Si,
\begin{equation}
n_{\mathrm{SiN},c}(r)=\frac{1}{N_{\mathrm{Si},c}}
\sum_{i\in\mathrm{Si}}\sum_{j\in\mathrm N,\,\mathrm{images}}
\ind\{r_{ij}\leq r\}.
\label{eq:running}
\end{equation}
The first-shell plateau and $n_{\mathrm{SiN},c}(\SI{2.33}{\angstrom})$ check the coordination obtained from the bond graph. Peak positions are reported at the resolution of the \SI{0.01}{\angstrom} histogram bins.

\subsection{Ring and density measures}
\begin{figure}[p]
\centering
\includegraphics[width=\textwidth,height=0.74\textheight,keepaspectratio]{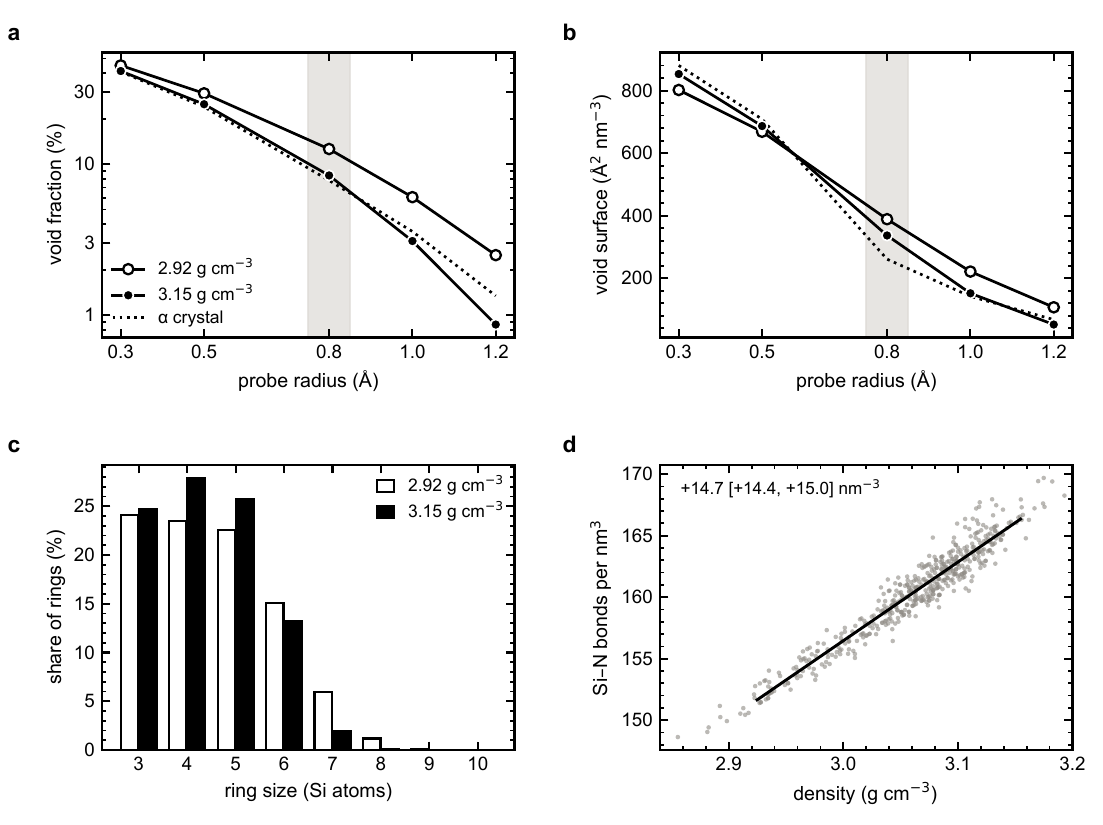}
\caption{\textbf{Probe sensitivity and network packing.} (a) Accessible void fraction and (b) specific internal surface against probe radius. Hollow and filled symbols are fitted values at 2.92 and \SI{3.15}{\gram\per\centi\metre\cubed}, respectively; dotted curves give crystalline $\alpha$-Si$_3$N$_4$ evaluated by the same clearance measure. Shading marks the \SI{0.8}{\angstrom} probe used in the main text. (c) Fitted primitive-ring fractions at the same densities, with ring size counted in Si atoms. (d) Si--N bond number density; the annotation gives its fitted change from 2.92 to \SI{3.15}{\gram\per\centi\metre\cubed} and the 95\% cell-bootstrap interval.}
\label{fig:S3}
\end{figure}

Primitive-ring populations were obtained from the per-cell structural analysis. For ring counts $R_c(s)$ at ring-size coordinate $s$, the per-cell fraction and mean are
\begin{equation}
p_c(s)=\frac{R_c(s)}{\sum_tR_c(t)},\qquad
\bar s_c=\sum_ssp_c(s).
\label{eq:ring}
\end{equation}
Ring size $s$ counts Si atoms in the alternating Si--N cycle. Figure~\ref{fig:S3}c reports the per-cell ring fractions evaluated at the reference densities of Section~\ref{sec:stats}.

Mass density is calculated from composition and relaxed volume,
\begin{equation}
\rho_c=\frac{120M_{\mathrm{Si}}+160M_{\mathrm N}}{N_{\mathrm A}V_c},
\label{eq:density}
\end{equation}
with the appropriate conversion from \AA$^3$ to cm$^3$. Bond and ring number densities are $1000N_{\mathrm{bond},c}/V_c$ and $1000N_{\mathrm{ring},c}/V_c$ when reported per \SI{1000}{\angstrom\cubed}. The volume-normalised counts quantify network packing alongside the per-cell connectivity measures.

\section{Void geometry and structural precursor fingerprint}
\label{sec:voids}

\subsection{Periodic covalent clearance}
The signed clearance at a grid point $\mathbf x$ is
\begin{equation}
C_c(\mathbf x)=\min_{i,\mathbf n}\left[
\left|\mathbf x-\mathbf r_i-H_c\mathbf n\right|-r_i^{\mathrm{cov}}\right],
\label{eq:clearance}
\end{equation}
where $H_c$ is the cell matrix and $\mathbf n$ indexes periodic images. The implementation uses the primary cell and its 26 neighbouring images with a nearest-neighbour search. Cordero radii are \SI{1.11}{\angstrom} for Si and \SI{0.71}{\angstrom} for N.\cite{Cordero2008Covalent} The field is sampled at approximately \SI{0.15}{\angstrom} spacing. Positive clearance denotes space outside the covalent spheres.

For probe radius $\rp$, the geometrically accessible set, its volume and volume fraction are
\begin{equation}
\Omega_c(\rp)=\{\mathbf x\in V_c:C_c(\mathbf x)\geq\rp\},\qquad
V_{\mathrm v,c}(\rp)=\int_{V_c}\ind\{C_c(\mathbf x)\geq\rp\}\,d\mathbf x,
\qquad \phi_c(\rp)=\frac{V_{\mathrm v,c}(\rp)}{V_c}.
\label{eq:voidvolume}
\end{equation}
The accessible volume includes all grid regions that accommodate the probe, including isolated cavities. The void surface is the isosurface $C_c(\mathbf x)=\rp$. Marching cubes is applied after tiling the periodic field twice along each lattice direction. Triangles are assigned to the primary cell by their centroids, closing boundary-spanning surfaces and counting each triangle once. Specific surface is $1000A_c(\rp)/V_c$ in \AA$^2$ per \SI{1000}{\angstrom\cubed}.

The main probe radius is \SI{0.8}{\angstrom}. A sweep over 0.3, 0.5, 0.8, 1.0 and \SI{1.2}{\angstrom} measures the dependence of accessible void fraction and specific internal surface on probe size (Fig.~\ref{fig:S3}a,b). Crystalline $\alpha$-Si$_3$N$_4$ was evaluated with the same clearance field and probe radii as the interstitial reference. At the main probe radius its accessible void fraction is 7.8\%. Independent Voronoi calculations provide the largest included-pore diameter $D_i$ and free-sphere diameter $D_f$. Candidate-site free volumes and cell free-volume fractions were obtained from the atom-resolved free-volume analysis.

\subsection{Candidates, compensation and local strain}
All three- and four-coordinate Si sites in the neutral structure were included as candidate environments. A three-coordinate Si bonded to a two-coordinate N is labelled locally compensated; a three-coordinate Si without such a neighbour is labelled uncapped. Gap-state hosting and subsequent capture were evaluated for each class. Bond-length distributions and the rank of the Si--N contact at the candidate site characterise the local geometry associated with compensation.

Each candidate is described by its local free volume, distance to the nearest cavity and bond-and-angle strain score,
\begin{equation}
S_i=\frac{\Delta d_i-\mu_{\mathrm{ref},d}}{s_{\mathrm{ref},d}}
+\frac{\Delta\theta_i-\mu_{\mathrm{ref},\theta}}{s_{\mathrm{ref},\theta}}.
\label{eq:strain}
\end{equation}
The reference means and standard deviations were computed from candidates pooled across both sources and applied unchanged to all retained cells for comparisons across density and route prediction. Realised trap sites were then associated with their neutral descriptors.

For an initially four-coordinate Si, the neutral Si--N distances are ordered as $d_1^0\leq\cdots\leq d_4^0$. The fourth distance $d_4^0$ characterises the tensile-strain tail. The three per-cell predictor inputs are
\begin{equation}
\mathbf x_c=\left(N_{\mathrm{free},c},\ N_{\mathrm{tail},c},\ S_{\max,c}\right),
\quad
N_{\mathrm{tail},c}=\sum_{i:z_i^0=4}\ind\{d_{4,i}^0>\SI{1.90}{\angstrom}\},
\quad S_{\max,c}=\max_{i:z_i^0=4}S_i.
\label{eq:supplies}
\end{equation}
$N_{\mathrm{free}}$ counts uncapped three-coordinate Si and is zero in cells without such sites. Charged-state connectivity was evaluated using the bond graph in Eq.~\eqref{eq:graph}.

\subsection{Descriptor contrasts and route prediction}
\label{sec:prediction}
For descriptor $x_j$, main-text Fig.~1a reports
\begin{equation}
z_{\mathrm{shift},j}=\frac{\bar x_{L,j}-\bar x_{H,j}}{s_{H,j}},
\label{eq:effectsd}
\end{equation}
where $L$ and $H$ denote the least dense and densest thirds of the 499 cells. Error bars are 95\% intervals calculated from the pooled standard error of the mean difference and expressed in the same reference units. This structural contrast is evaluated between density thirds; the density-trend comparisons use the reference values defined in Section~\ref{sec:stats}.

The three cell-level precursor descriptors are standardised using the training-half means and standard deviations. For standardised inputs $\widetilde{\mathbf x}_c$, the multinomial predictor is
\begin{equation}
p_m(\mathbf x_c)=
\frac{\exp(a_m+\boldsymbol\beta_m^{\mathsf T}\widetilde{\mathbf x}_c)}
{\sum_{r\in\{K,iK,\mathrm{pol}\}}\exp(a_r+\boldsymbol\beta_r^{\mathsf T}\widetilde{\mathbf x}_c)}.
\label{eq:softmax}
\end{equation}
Coefficients were fitted by $L_2$-regularised gradient descent on a random half of the 496 routed cells. The fitted transformation and coefficients were frozen and applied to the remaining cells. Overall predicted route fractions are the mean cell probabilities, $\widehat\pi_m=n^{-1}\sum_c p_m(\mathbf x_c)$. Predicted density-dependent fractions are evaluated from the held-out probabilities using the same density-trend procedure as the observed route indicators. Main-text Fig.~4e reports the overall fractions, and Fig.~\ref{fig:S9}d reports the density-resolved comparison.

Intervals for predicted route fractions are 95\% percentile intervals from 400 whole-cell bootstrap resamples. Repeating the random-half split 200 times gives 60\% correct held-out route assignments, compared with 41\% using route fractions alone (Fig.~\ref{fig:S9}c). Source-resolved route and candidate comparisons are retained in Fig.~\ref{fig:S1}e,f.

Figure~\ref{fig:S9}b reports coefficient contrasts relative to polaronic trapping, $\boldsymbol\beta_m-\boldsymbol\beta_{\mathrm{pol}}$. Per standard deviation of the input, the uncapped-Si coefficient for K-centre capture is $+2.4\,[2.0,3.0]$ and the tensile-tail coefficient for induced-K formation is $+0.43\,[0.25,0.65]$. Adding accessible void fraction gives a held-out log loss of 0.769, compared with 0.763 for the three-input predictor. The two-strain-input comparison omits the uncapped-Si supply (Fig.~\ref{fig:S9}d).

\section{Electronic precursors, capture and structural relaxation}
\label{sec:electronic}

\subsection{State participation and band character}
Electronic participation is reported as an effective number of atoms. For atom-resolved non-negative state weights $w_{ia}$,
\begin{equation}
I_i=\frac{\sum_a w_{ia}^2}{(\sum_a w_{ia})^2},\qquad
N_{\mathrm{eff},i}=I_i^{-1}.
\label{eq:ipr}
\end{equation}
The occupied and empty state profiles provide the atom-count localisation measures used to identify the delocalisation onsets. Real-space effective-atom participation describes the precursor-to-trap change in the main text, and molecular-orbital participation provides an independent comparison. The coefficient-space diagnostic used in the original source is $\sum_\mu|c_{i\mu}|^4/(\sum_\mu|c_{i\mu}|^2)^2$.\cite{Hueckmann2024Intrinsic}

Species character was calculated from the Si and N projections of each state as $f_{\mathrm{Si}}=P_{\mathrm{Si}}/(P_{\mathrm{Si}}+P_{\mathrm N})$ and summarised over the valence and conduction windows. This measures the chemical character of the states surrounding the gap.

\subsection{Electrostatic potential at sites and over precursor states}
\label{sec:potential}
The neutral electrostatic potential energy of an electron was averaged within \SI{1.0}{\angstrom} of each nucleus. For Si site $i$, let $V_i$ denote this average and $\langle V\rangle_{\mathrm{Si}}$ its mean over the 120 Si sites of the cell. The well depth and confinement are
\begin{equation}
D_i^V=\langle V\rangle_{\mathrm{Si}}-V_i,\qquad
Q_i^V=\langle V\rangle_{1.4<r_i<3.0\,\text{\AA}}-V_i.
\label{eq:sitepotential}
\end{equation}
Positive depth denotes a site below the cell-average Si potential. Confinement measures the potential rise from the site to the surrounding 1.4--\SI{3.0}{\angstrom} shell. For Si-centred trapping, site depths were ranked within each cell; Fig.~\ref{fig:S6}d reports the fraction of trapping sites among the $k$ deepest Si wells.

For the state-resolved potential (Fig.~\ref{fig:S5}), grid points were ordered by the added-electron density at fixed nuclei. Successive shells between isodensity surfaces each contain 0.5\% of the electron. The potential is the density-weighted shell mean, referenced to the surroundings of an average atom of the same species. An enclosed volume $V$ is represented by the equivalent radius $r_{\mathrm{eq}}=(3V/4\pi)^{1/3}$. The precursor and relaxed-state renderings enclose 50\% of the electron density; the circle in the precursor rendering marks the trap-centred radius containing the same fraction.

\subsection{Mobility-gap construction and threshold sensitivity}
Mobility edges and spectral gaps were evaluated from the charge-corrected eigenvalues. Let $N=280$ and let $\eta$ denote the required participating fraction of a cell. The conduction onset is the lowest empty state satisfying $\Neff\geq\eta N$; the valence onset is the highest occupied state satisfying the same criterion:
\begin{equation}
E_{\mathrm c}^{(\eta)}=\min_{i\in\mathrm{empty}:N_{\mathrm{eff},i}\geq\eta N}\epsilon_i^{\mathrm{corr}},
\qquad
E_{\mathrm v}^{(\eta)}=\max_{i\in\mathrm{occupied}:N_{\mathrm{eff},i}\geq\eta N}\epsilon_i^{\mathrm{corr}},
\qquad
E_{\mathrm g,\mathrm{mob}}^{(\eta)}=E_{\mathrm c}^{(\eta)}-E_{\mathrm v}^{(\eta)}.
\label{eq:edges}
\end{equation}
State index $k=0$ denotes the HOMO and $k=1$ the LUMO. The localisation profiles cover approximately 50 states on either side of the gap. Valence-state energies were obtained from the corresponding projected-density-of-states output and included in the corrected spectral analysis.

\begin{figure}[p]
\centering
\includegraphics[width=\textwidth,height=0.74\textheight,keepaspectratio]{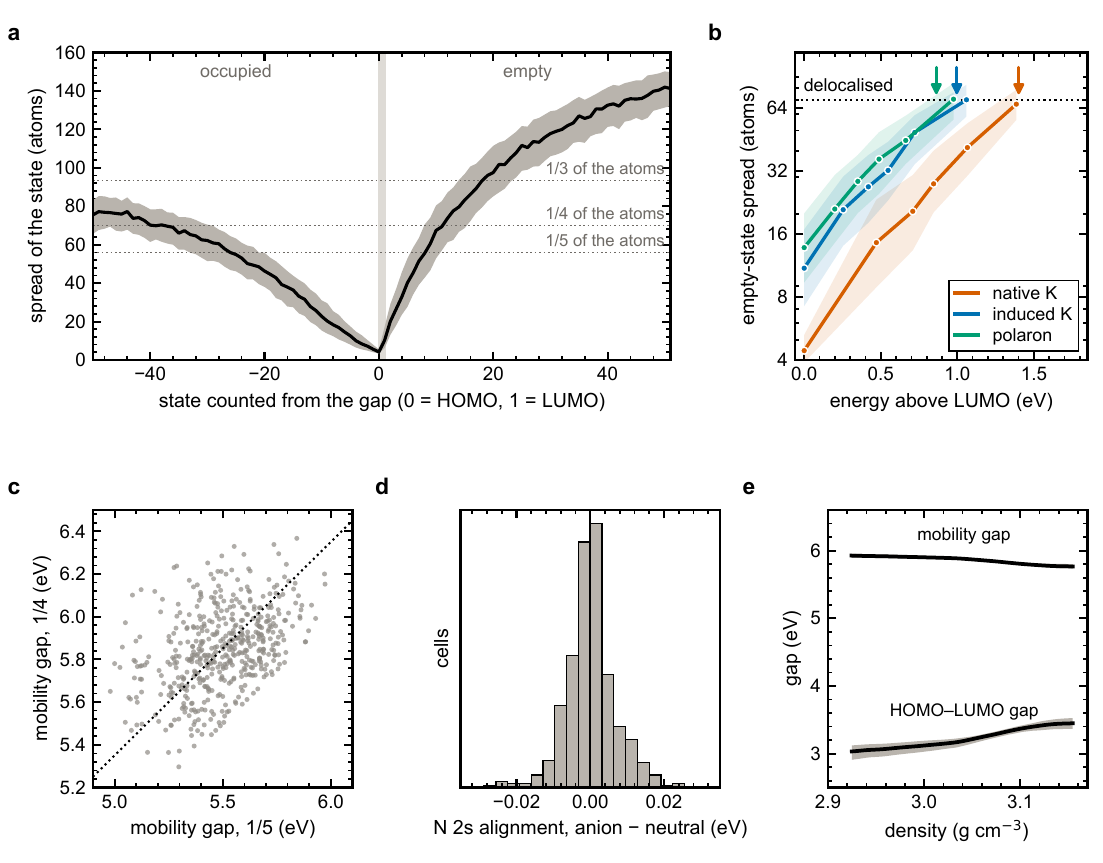}
\caption{\textbf{Mobility-edge definition and spectral alignment.} (a) Median effective-atom participation against state index ($k=0$ HOMO, $k=1$ LUMO), with the interquartile range shaded. Horizontal lines mark $N/5$, $N/4$ and $N/3$. (b) Empty-state participation above the LUMO, resolved by trapping route; arrows mark the $N/4$ delocalisation onset. (c) Per-cell mobility gaps at $N/5$ and $N/4$; the dotted line marks the mean offset of \SI{0.35}{\electronvolt}. (d) N $2s$ reference shifts between neutral and anion spectra. (e) Density-dependent HOMO--LUMO and mobility gaps, the latter evaluated using $N/4$. Spectral quantities use the corrected eigenvalues and alignment defined in Section~\ref{sec:corrected-levels}.}
\label{fig:S4}
\end{figure}

\begin{figure}[p]
\centering
\includegraphics[width=\textwidth,height=0.74\textheight,keepaspectratio]{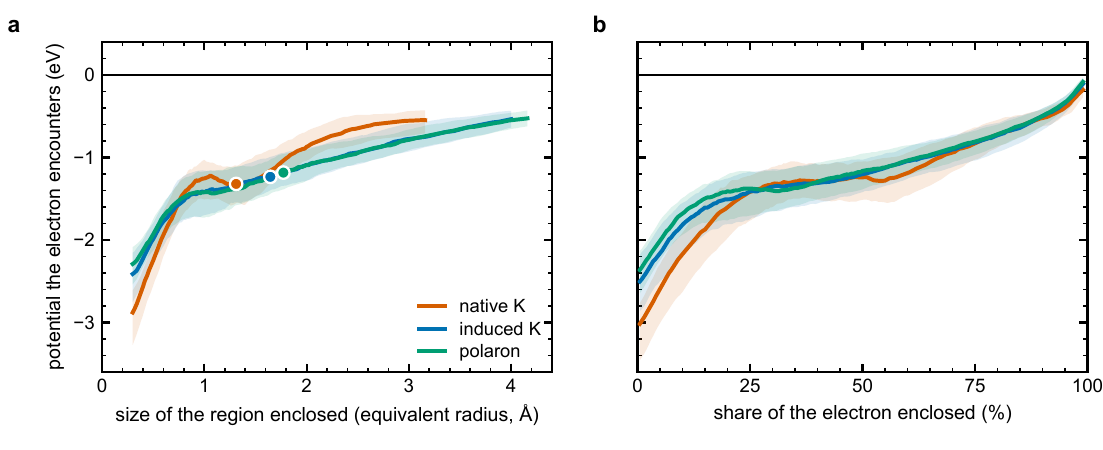}
\caption{\textbf{Electrostatic potential sampled by the precursor electron.} Route medians (lines) and interquartile ranges (shading) of the density-weighted neutral potential in successive isodensity shells, each containing 0.5\% of the added electron before ionic relaxation. (a) Potential against the equivalent radius of the enclosed region, shown to 90\% enclosed charge; circles mark 50\%. (b) Potential against enclosed-electron fraction. The potential reference and shell construction are defined in Section~\ref{sec:potential}.}
\label{fig:S5}
\end{figure}

\begin{figure}[p]
\centering
\includegraphics[width=\textwidth,height=0.74\textheight,keepaspectratio]{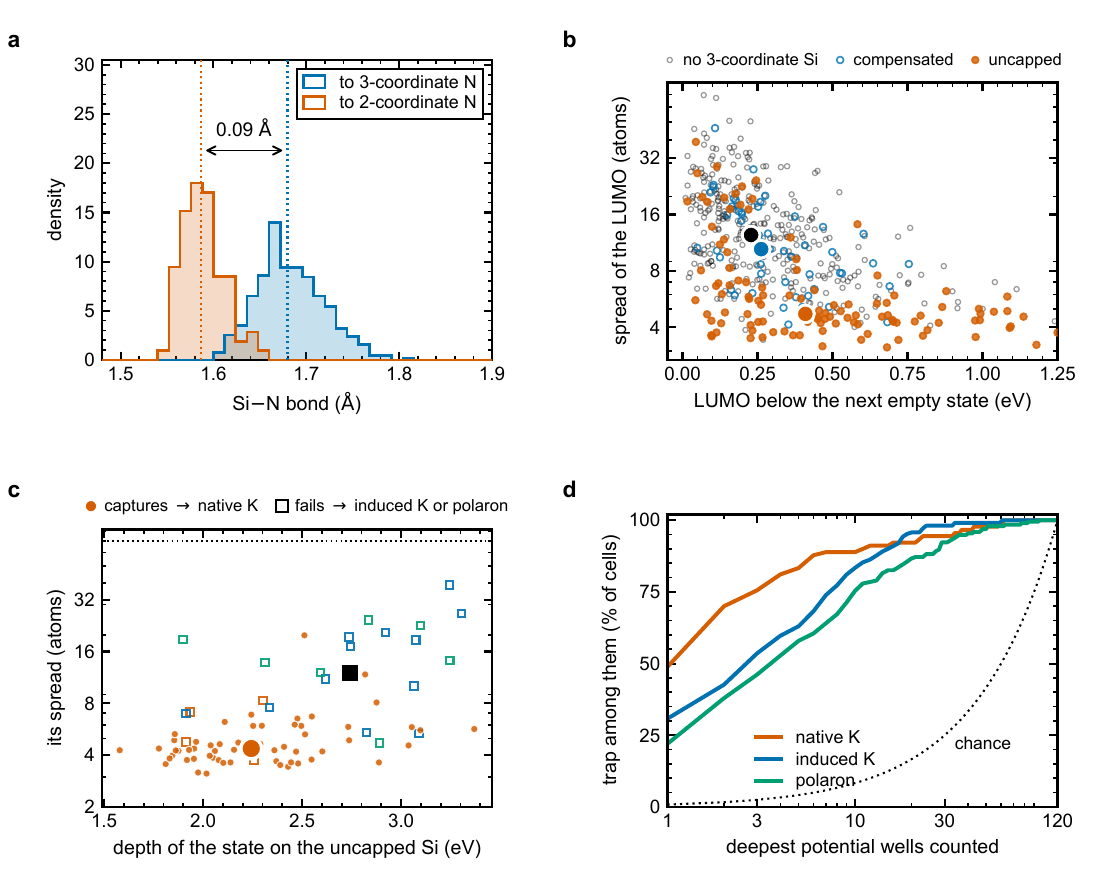}
\caption{\textbf{Local compensation and competition for capture.} (a) Si--N distances at three-coordinate Si sites bonded to two- or three-coordinate N; the distribution centres differ by \SI{0.09}{\angstrom}. (b) LUMO detachment from the next empty state and effective-atom spread for cells containing no three-coordinate Si, compensated-only sites or uncapped candidates. Large symbols are medians. (c) Candidate-state depth and spread for the source of Ref.~\cite{Cottom2026Vitriflow}; filled circles capture at the candidate and hollow squares follow induced-K or polaronic trapping elsewhere. (d) Fraction of trapping atoms among the $k$ deepest Si-site potential wells, resolved by route; the dotted curve is the uniform-site reference $k/120$.}
\label{fig:S6}
\end{figure}

The mobility gaps and trap depths use $\eta=1/4$, corresponding to participation over at least 70 atoms and an atom-weight IPR of $I\leq4/N$. Threshold sensitivity is shown for $\eta=1/5$ and $1/3$ in Fig.~\ref{fig:S4}a,c. The $N/3$ occupied-state crossing generally lies beyond the 50-state window.

Figure~\ref{fig:S4}c compares the per-cell mobility gaps obtained at $N/5$ and $N/4$. The change of threshold shifts the mean gap by approximately \SI{0.35}{\electronvolt}; the per-cell Spearman correlation is 0.42. The HOMO--LUMO gap was calculated from the corrected highest occupied and lowest empty eigenvalues, including localised gap and band-tail states.

\subsection{Charge-corrected eigenvalues, spectral alignment and level depths}
\label{sec:corrected-levels}
\begin{figure}[p]
\centering
\includegraphics[width=\textwidth,height=0.74\textheight,keepaspectratio]{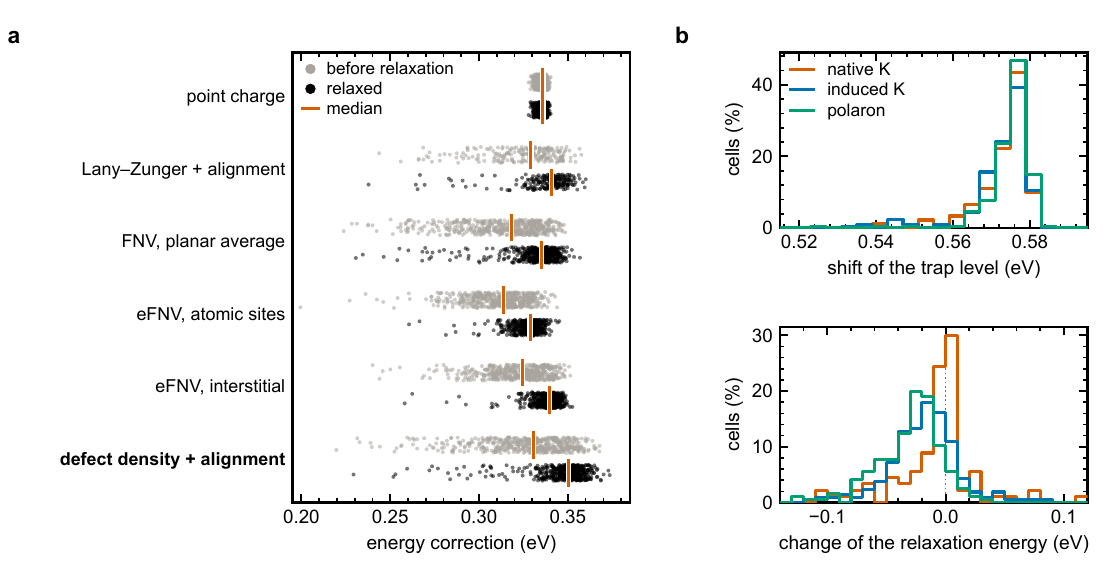}
\caption{\textbf{Finite-size corrections to total energies and trap eigenvalues.} (a) Total-energy corrections for vertical and relaxed $q=-1$ cells under six electrostatic treatments. Ticks mark medians; bold identifies the defect-density image-interaction plus charging-alignment treatment used in the analysis, with $\varepsilon=4.2$. (b) Route-resolved distributions of the trap-eigenvalue shift (upper) and the correction-induced change in positive relaxation energy, $E_{\mathrm{relax}}^{\mathrm{corr}}-E_{\mathrm{relax}}^{\mathrm{PBC}}$ (lower).}
\label{fig:S7}
\end{figure}

Spectral comparisons use finite-size-corrected Kohn--Sham eigenvalues\cite{Naik2018CoFFEE}. The applied correction combines the image interaction of the defect charge density with charging alignment~\cite{Durrant2018ImageCharge}, using a dielectric constant $\varepsilon=4.2$. Figure~\ref{fig:S7}a compares its total-energy corrections with point-charge, Lany--Zunger, planar FNV and atomic-site/interstitial eFNV treatments\cite{Lany2009,Freysoldt2009FNV,Kumagai2014FiniteSize}; the median corrections differ by a few hundredths of an electronvolt.  The eigenvalue correction follows the identified trap orbital. The correction changes each route's median positive relaxation energy by no more than \SI{0.02}{\electronvolt} (Fig.~\ref{fig:S7}b). For state $i$ in charge state $q$ at the geometry being analysed, $\delta\epsilon_i^q$ denotes the applied eigenvalue correction. The corrected spectra are expressed on a common energy scale using the N $2s$ reference:
\begin{equation}
\begin{aligned}
\epsilon_i^{q,\mathrm{corr}}
  &=\epsilon_i^{q,\mathrm{PBC}}+\delta\epsilon_i^q,\\
\Delta_{2s}
  &=\bar\epsilon_{2s}^{0,\mathrm{corr}}
    -\bar\epsilon_{2s}^{-,\mathrm{corr}},\\
\widetilde\epsilon_i^{-,\mathrm{corr}}
  &=\epsilon_i^{-,\mathrm{corr}}+\Delta_{2s}.
\end{aligned}
\label{eq:alignment}
\end{equation}
The superscripts $0$ and $-$ identify the neutral and singly charged calculations, respectively; the neutral spectrum defines the alignment zero. The N $2s$ alignment shift is $0.000\pm\SI{0.007}{\electronvolt}$ (mean $\pm$ standard deviation; Fig.~\ref{fig:S4}d).

Trap depth was calculated from the corrected trap eigenvalue and corrected conduction reference in the same cell and charge state,
\[
D_t^{q,\mathrm{corr}}
  =E_{\mathrm c}^{q,\mathrm{corr}}-\epsilon_t^{q,\mathrm{corr}}.
\]
The paired depth comparison tracks the change from the neutral precursor to the relaxed charged state on the corrected spectral scale. The independently determined neutral and charged mobility edges agree within \SI{0.01}{\electronvolt}, placing the precursor-to-trap deepening on a common reference scale.

\subsection{Charge-corrected relaxation energy}
\label{sec:corrected-energy}
The charged-state relaxation energy was calculated from the corrected total energies of the vertical and relaxed $q=-1$ configurations, using the defect-density image interaction and charging alignment with $\varepsilon=4.2$ at both endpoints. The vertical energy is obtained after electronic convergence at the neutral equilibrium geometry, before ionic motion. The relaxed energy follows ionic optimisation of the charged cell. Both endpoints contain one excess electron and have identical composition and cell dimensions.

For endpoint $\alpha\in\{\mathrm{vertical},\mathrm{relaxed}\}$, let $E_{\alpha}^{-,\mathrm{PBC}}$ be the periodic-supercell total energy and $C^-_{\alpha}$ its additive electrostatic finite-size correction. The corrected energies and their difference are
\begin{equation}
\begin{aligned}
E_{\alpha}^{-,\mathrm{corr}}
  &=E_{\alpha}^{-,\mathrm{PBC}}+C^-_{\alpha},\\
\Delta E_{\mathrm{rel}}^{\mathrm{corr}}
  &=E_{\mathrm{relaxed}}^{-,\mathrm{corr}}
    -E_{\mathrm{vertical}}^{-,\mathrm{corr}}\\
  &=\bigl(E_{\mathrm{relaxed}}^{-,\mathrm{PBC}}
    -E_{\mathrm{vertical}}^{-,\mathrm{PBC}}\bigr)
    +C^-_{\mathrm{relaxed}}-C^-_{\mathrm{vertical}}.
\end{aligned}
\label{eq:relaxenergy}
\end{equation}
The correction is applied to each charged endpoint, retaining the differential contribution $C^-_{\mathrm{relaxed}}-C^-_{\mathrm{vertical}}$. A negative $\Delta E_{\mathrm{rel}}^{\mathrm{corr}}$ denotes favourable relaxation. The positive stabilisation energy, $E_{\mathrm{relax}}=-\Delta E_{\mathrm{rel}}^{\mathrm{corr}}$, is reported in main-text Fig.~4b,d. Figure~\ref{fig:S7}b shows the correction-induced change in this quantity, $E_{\mathrm{relax}}^{\mathrm{corr}}-E_{\mathrm{relax}}^{\mathrm{PBC}}$.

\subsection{Candidate activation and capture}
\label{sec:activation}
Gap-state hosting was evaluated for each neutral three-coordinate Si and grouped by local compensation. Whole-cell comparisons distinguish cells with no three-coordinate Si, compensated-only sites and at least one uncapped candidate. Candidate availability, gap-state hosting and capture were recorded separately. Candidate-resolved depth and participation were compared in the ref. \cite{Cottom2026Vitriflow} source; lowest-empty-state selection was evaluated across the retained population. Depths and state ordering use the corrected eigenvalues. The LUMO detachment in Fig.~\ref{fig:S6}b is its separation from the next empty state, $\epsilon_{\mathrm{LUMO}+1}^{0,\mathrm{corr}}-\epsilon_{\mathrm{LUMO}}^{0,\mathrm{corr}}$.

Vertical spin weight was evaluated on the final trapping atom, and its rank among the atomic spin weights was recorded in every retained cell. Relocation compares the vertical and relaxed localisation centres. Capture and relocation summaries are reported by route, with source-resolved checks in Section~\ref{sec:population}.

Trapping routes were assigned from the paired neutral and charged calculations. Native K denotes capture at a K centre already present in the neutral structure. Induced-K (iK) denotes formation of a three-coordinate Si through Si--N bond detachment during charged relaxation. A polaron retains the local Si--N shell as the network distorts. Si--Si reconstruction is flagged alongside the route assignment, and all three such cells remain in the routed population. Native and induced describe K-centre provenance; all responses are intrinsic to the amorphous network.

\subsection{Displacement shells and bond response}
\begin{figure}[p]
\centering
\includegraphics[width=\textwidth,height=0.74\textheight,keepaspectratio]{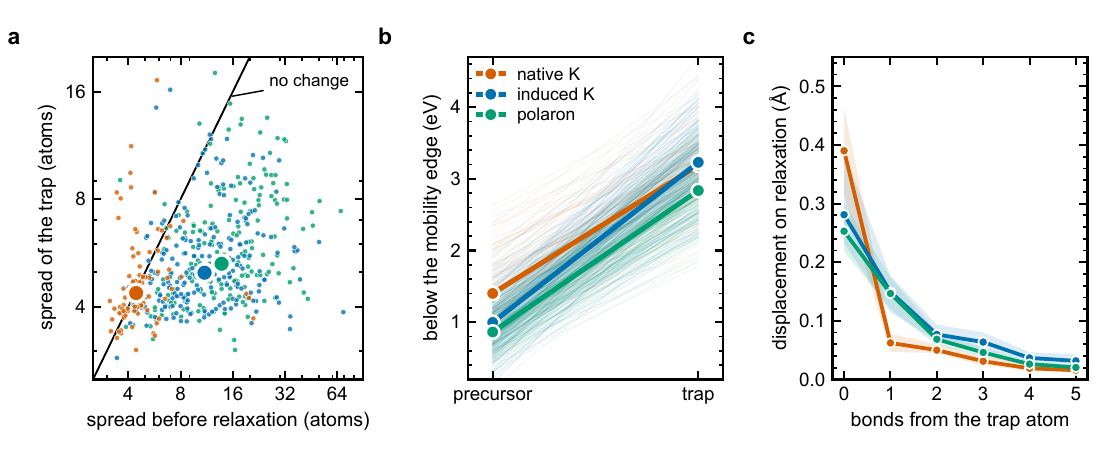}
\caption{\textbf{Localisation and ionic response during relaxation.} (a) Effective-atom spread of the precursor and relaxed trap; large symbols are route medians and the diagonal marks unchanged participation. (b) Paired precursor and relaxed-trap depths on the corrected mobility-edge reference; fine lines represent individual cells and thick lines route medians. (c) Mean ionic displacement within each graph shell of a cell, summarised by the route median and interquartile range. Shell zero is the trapping atom; shell one contains its bonded neighbours. Displacements are reported in \AA.}
\label{fig:S8}
\end{figure}

\begin{figure}[p]
\centering
\includegraphics[width=\textwidth,height=0.74\textheight,keepaspectratio]{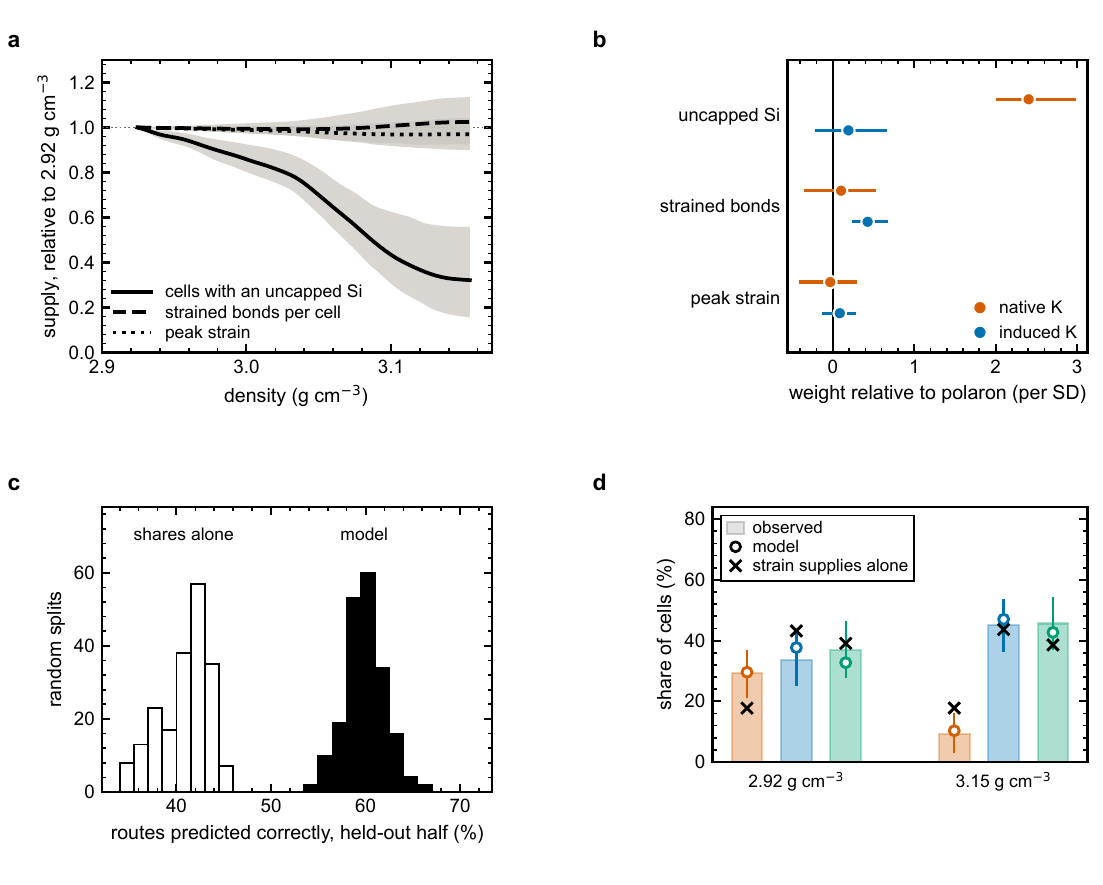}
\caption{\textbf{Neutral precursor populations and route prediction.} (a) Density dependence of the fraction of cells with an uncapped Si (solid), mean tensile-tail count (dashed) and mean peak strain (dotted), normalised to their fitted values at \SI{2.92}{\gram\per\centi\metre\cubed}. Shading gives 95\% cell-bootstrap intervals. The model uses the uncapped-Si count defined in Eq.~\eqref{eq:supplies}. (b) Multinomial coefficient contrasts relative to polaronic trapping, per standard deviation of each input, with 95\% bootstrap intervals. (c) Correct held-out route assignments over 200 random-half splits, compared with the route-fraction baseline. (d) Observed bars and predicted circles at 2.92 and \SI{3.15}{\gram\per\centi\metre\cubed}, evaluated from held-out cells; error bars are 95\% cell-bootstrap intervals. Crosses use only the two strain inputs. The density estimates follow Section~\ref{sec:stats}.}
\label{fig:S9}
\end{figure}

Atomic identities were retained between the neutral and relaxed charged geometries. Periodic displacement magnitudes were grouped by bond-graph distance from the final trapping atom. For shell $h$ in cell $c$, the mean displacement is
\begin{equation}
\bar u_c(h)=\frac{1}{|\mathcal S_{c,h}|}
\sum_{i\in\mathcal S_{c,h}}|\Delta\mathbf r_i|,
\label{eq:shelldisplacement}
\end{equation}
where $\mathcal S_{c,0}$ contains the trap atom and $\mathcal S_{c,1}$ its bonded neighbours. Figure~\ref{fig:S8}c reports the route median and interquartile range of these per-cell shell means, in \AA, for shells zero to five.

Displacement participation from the per-cell relaxation analysis quantifies the number of atoms participating in the structural response. To resolve the local bond response, neutral neighbours were ranked by Si--N length and the same atom pairs followed after relaxation:
\begin{equation}
\Delta d_j=d_j(R^-)-d_j(R^0),\qquad j=1,\ldots,4.
\label{eq:bondchange}
\end{equation}
The fourth-bond plot relates the neutral tensile-strain proxy $d_4^0$ to the charged-state extension $\Delta d_4$, grouped by the independently assigned trapping route. This tests how the initial tensile distortion predicts polaronic relaxation or induced-K formation.

\section{Statistical analysis}
\label{sec:stats}

\subsection{Source summaries and density contrasts}
The analysis covers 2.86--\SI{3.19}{\gram\per\centi\metre\cubed}. Source means and distributions in Figs.~\ref{fig:S1} and \ref{fig:S2} summarise the two published sources separately. Main-text Fig.~1a compares the least dense and densest thirds of the combined population using Eq.~\eqref{eq:effectsd}.

Density-dependent changes are evaluated at $\rho_L=2.92$ and $\rho_H=\SI{3.15}{\gram\per\centi\metre\cubed}$, the 2.5th and 97.5th percentiles of the retained density distribution. These reference values define the low- and high-density estimates in the main-text figures, Table~\ref{tab:routes} and the probe and predictor comparisons. Source means, density-third contrasts and fitted reference-density values are retained as separate summaries.

\subsection{Density trends and uncertainty}
Cells were ordered by density. Smooth trends were calculated using local-linear fits with a Gaussian window of 200 cells, selected by leave-one-out prediction of the trapping route. The fitted trend was evaluated at $\rho_L$ and $\rho_H$. Route fractions use cell-level route indicators, and predicted fractions use the corresponding held-out model probabilities. Density-dependent changes refer to the difference between these fitted values.

Uncertainty is expressed as 95\% percentile intervals from 400 bootstrap resamples of whole cells, preserving the grouping of sites, bonds and electronic states within each cell. Route fractions compared by source over 2.96--\SI{3.02}{\gram\per\centi\metre\cubed} carry 95\% Wilson intervals (Fig.~\ref{fig:S1}e). Per-cell histograms use common bin edges and unit-area normalisation, with empirical cumulative distributions beneath the same axes. Bond-, angle- and site-level distributions describe the corresponding local environments.

Figure~\ref{fig:S9}a shows the fraction of cells containing at least one uncapped Si, the mean tensile-tail count and the mean peak strain, each divided by its fitted value at $\rho_L$. The first curve uses the indicator $\ind\{N_{\mathrm{free},c}>0\}$; the predictor itself uses the count $N_{\mathrm{free},c}$ from Eq.~\eqref{eq:supplies}. Model fitting, repeated random-half tests and intervals for predicted fractions are defined in Section~\ref{sec:prediction}.

\subsection{Within-route energy changes}
\label{sec:within}
Corrected trap depths and positive charged-state relaxation energies were fitted linearly against density within each route. For quantity $Y$ and route $m$,
\begin{equation}
\widehat Y_m(\rho)=a_m+b_m\rho,\qquad
\Delta Y_m=\widehat Y_m(\rho_H)-\widehat Y_m(\rho_L)
=b_m(\rho_H-\rho_L).
\label{eq:withincontrast}
\end{equation}
The reference-density estimates in main-text Fig.~4d and the differences in Table~\ref{tab:within} use the same fitted line and cell selection for each quantity and route. Intervals were obtained by refitting after whole-cell bootstrap resampling. Negative values denote lower energies at higher density.

\begin{table}[htbp]
\centering
\caption{Fitted changes in corrected trap depth and positive relaxation energy from 2.92 to \SI{3.15}{\gram\per\centi\metre\cubed}. Entries are higher-density minus lower-density values from the within-route linear fits, with 95\% cell-bootstrap intervals. The same fits supply the endpoint estimates in main-text Fig.~4d.}
\label{tab:within}
\small
\begin{tabular}{@{}lccc@{}}
\toprule
Quantity (eV) & K centre & Induced K & Polaron\\
\midrule
Trap depth & $-0.12\,[-0.51,0.28]$ & $-0.17\,[-0.45,0.11]$ & $-0.05\,[-0.33,0.22]$\\
Relaxation energy & $-0.14\,[-0.38,0.17]$ & $-0.06\,[-0.26,0.16]$ & $-0.03\,[-0.17,0.10]$\\
\bottomrule
\end{tabular}
\end{table}

\clearpage
\section{Data organisation and reproducibility}
\label{sec:records}
Per-cell tables were joined using the source identifier and canonical integer cell key. Each derived table has a machine-readable provenance record containing its sources, analysis method and hashes. Table~\ref{tab:data} links the supplementary figures and the within-route comparison to their analysis products.

\begin{longtable}{@{}>{\raggedright\arraybackslash}p{0.12\textwidth}>{\raggedright\arraybackslash}p{0.41\textwidth}>{\raggedright\arraybackslash}p{0.38\textwidth}@{}}
\caption{Analysis products associated with the supplementary figures and within-route energy comparison.}\label{tab:data}\\
\toprule
Item & Derived tables & Main inputs\\
\midrule
\endfirsthead
\toprule Item & Derived tables & Main inputs\\\midrule
\endhead
S1 & \texttt{si\_S1\_distributions}, \texttt{si\_S1\_routes\_in\_overlap}, \texttt{si\_S1\_rule\_by\_set} & Per-cell structural, void, band-edge and route tables; source identifiers.\\[4pt]
S2 & \texttt{pf\_atlas\_gr}, \texttt{pf\_atlas\_firstmin}, \texttt{pf\_atlas\_geomhist}, \texttt{pf\_atlas\_geomstat} & Periodic structures, common RDF binning and the \SI{2.33}{\angstrom} bond graph.\\[4pt]
S3 & \texttt{si\_S3\_probe\_sweep}, \texttt{pf\_crystal\_reference}, \texttt{hse\_structure\_per\_cell} & Periodic clearance, crystalline references, ring and bond counts.\\[4pt]
S4 & \texttt{pf\_rig\_cells}, \texttt{pf\_manifold\_flavour} & Wide IPR and corrected mobility-edge and band-reference data.\\[4pt]
S5 & \texttt{si\_S5\_state\_potential} & Neutral electrostatic potential and vertical added-electron density.\\[4pt]
S6 & \texttt{hse\_gap\_state\_sites}, \texttt{hse\_gap\_state\_bonds}, \texttt{pf\_competition}, \texttt{pf\_potential\_rank} & Candidate sites, bonds, electronic states and site potentials.\\[4pt]
S7 & \texttt{si\_S7\_correction} & Total-energy and trap-orbital corrections at the charged endpoints.\\[4pt]
S8 & \texttt{si\_S8\_displacement}, \texttt{pf\_fig4\_routes} & Precursor and relaxed states; ionic displacements by graph shell.\\[4pt]
S9 & \texttt{si\_S9\_supplies}, \texttt{si\_S9\_coefficients}, \texttt{si\_S9\_splits}, \texttt{pf\_fig4\_density} & Routes, neutral precursor descriptors and held-out predictions.\\[4pt]
Table~\ref{tab:within} & \texttt{pf\_fig4\_within\_route} & Corrected trap depths and relaxation energies, cell densities and route assignments.\\
\bottomrule
\end{longtable}

The source identifier and cell key connect each neutral structure, its electronic states and its charged response. Cell selections, route assignments and Si--Si reconstruction flags are retained with the derived data products.
\clearpage

\clearpage
\begingroup
\small
\setlength{\bibsep}{2pt}
\bibliographystyle{elsarticle-num}
\bibliography{references}
\endgroup